\documentclass[final,11pt]{article2}

\usepackage{graphicx}
\usepackage{amssymb}
\usepackage{rotating}
\usepackage{longtable}
\usepackage{epsfig}
\usepackage{txfonts}
\usepackage{natbib}

\newcommand\farcs{\mbox{$.\!\!^{\prime\prime}$}}
\newcommand\arcsec{\mbox{$~\!\!^{\prime\prime}$}}

\begin{document}

\title{VISTA Variables in the Via Lactea (VVV): The~public ESO near-IR
  variability survey of the Milky~Way}

\maketitle

\begin{center}
\author{
D.~Minniti$^{1,2,\dagger}$, 
P.~W.~Lucas$^{3, \dagger}$, 
J.~P.~Emerson$^{4, \dagger}$, 
R.~K.~Saito$^{1}$, 
M.~Hempel$^{1}$, 
P.~Pietrukowicz$^{1,5}$, 
A.~V.~Ahumada$^{6,7,8}$,
M.~V.~Alonso$^{6}$, 
J.~Alonso-Garc\'{i}a$^{9}$,  
J.~I.~Arias$^{10}$,  
R.~M.~Bandyopadhyay$^{11}$,
R.~H.~Barb\'a$^{10}$,  
B.~Barbuy$^{12}$,      
L.~R.~Bedin$^{13}$,           
E.~Bica$^{14}$,               
J.~Borissova$^{15}$,          
L.~Bronfman$^{16}$, 
G.~Carraro$^{7}$,        
M.~Catelan$^{1}$,            
J.~J.~Clari\'a$^{6}$,        
N.~Cross$^{17}$,              
R.~de~Grijs$^{18,19}$,           
I.~D\'{e}k\'{a}ny$^{20}$,             
J.~E.~Drew$^{3,21}$,            
C.~Fari\~na$^{22}$,    
C.~Feinstein$^{22}$,       
E.~Fern\'andez~Laj\'us$^{22}$,
R.~C.~Gamen$^{10}$,           
D.~Geisler$^{23}$,            
W.~Gieren$^{23}$,             
B.~Goldman$^{24}$,            
O.~A.~Gonzalez$^{25}$,         
G.~Gunthardt$^{10}$,          
S.~Gurovich$^{6}$,           
N.~C.~Hambly$^{17}$,          
M.~J.~Irwin$^{26}$,           
V.~D.~Ivanov$^{7}$,          
A.~Jord\'an$^{1}$,           
E.~Kerins$^{27}$,             
K.~Kinemuchi$^{10,23}$,          
R.~Kurtev$^{15}$,             
M.~L\'opez-Corredoira$^{28}$, 
T.~Maccarone$^{29}$,          
N.~Masetti$^{30}$,            
D.~Merlo$^{6}$,              
M.~Messineo$^{31,32}$, 
I.~F.~Mirabel$^{33,34}$,         
L.~Monaco$^{7}$,             
L.~Morelli$^{35}$,            
N.~Padilla$^{1}$,
T.~Palma$^{6}$,
M.~C.~Parisi$^{6}$,          
G.~Pignata$^{36}$,            
M.~Rejkuba$^{25}$,            
A.~Roman-Lopes$^{10}$,        
S.~E.~Sale$^{21}$,             
M.~R.~Schreiber$^{15}$,       
A.~C.~Schr\"oder$^{37,38}$,      
M.~Smith$^{39}$,              
L.~Sodr\'e~Jr.$^{12}$,        
M.~Soto$^{10}$,               
M.~Tamura$^{40}$,             
C.~Tappert$^{1}$,            
M.~A.~Thompson$^{3}$,         
I.~Toledo$^{1}$
\&  
M.~Zoccali$^{1}$
}    
\end{center}           

\begin{center}
{\it (Affiliations can be found after the references)}\\
\vskip 0.5em
$\dagger${\it Email addresses}: dante@astro.puc.cl~(D.~Minniti),
  P.W.Lucas@herts.ac.uk~(P.~W.~Lucas), j.p.emerson@qmul.ac.uk~(J.~P.~Emerson)
\end{center}
\vskip 1em

\begin{abstract}
We describe the public ESO near-IR variability survey (VVV) scanning the Milky
Way bulge and an adjacent section of the mid-plane where star formation
activity is high. The survey will take 1929~hours of observations with the
4-metre VISTA telescope during five years ($2010-2014$), covering $\sim10^9$
point sources across an area of 520~deg$^2$, including 33 known globular
clusters and $\sim$$350$ open clusters. The final product will be a deep
near-IR atlas in five passbands ($0.9-2.5~\mu$m) and a catalogue of more than
$10^{6}$ variable point sources. Unlike single-epoch surveys that, in most
cases, only produce 2-D maps, the VVV variable star survey will enable the
construction of a 3-D map of the surveyed region using well-understood distance
indicators such as RR~Lyrae stars, and Cepheids. It will yield important
information on the ages of the populations. The observations will be combined
with data from MACHO, OGLE, EROS, VST, Spitzer, HST, Chandra, INTEGRAL, WISE,
Fermi LAT, XMM-Newton, GAIA and ALMA for a complete understanding of the
variable sources in the inner Milky Way. This public survey will provide data
available to the whole community and therefore will enable further studies of
the history of the Milky Way, its globular cluster evolution, and the
population census of the Galactic Bulge and center, as well as the
investigations of the star forming regions in the disk. The combined variable
star catalogues will have important implications for theoretical
investigations of pulsation properties of stars.
\end{abstract}

\noindent{{\it Key words}: Surveys, Stars: variables: general, Galaxy: bulge,
  Galaxy: disk }

\section{Introduction}

The bulk of the stars, gas and dust in the Milky Way are confined to its bulge
and plane.  As a result, in these directions, the extinction and crowding are
high, making any study of the inner structure of the Galaxy difficult. Knowing
how the stellar populations are distributed within the Galaxy is essential for
such studies and hence the main goal of the described survey.  Traditional
distance indicators have been used with various success in the past. The
approach was to concentrate on clear ``windows'', where optical surveys can be
carried out (e.g., MACHO, OGLE, EROS). In this paper we describe the VISTA
Variables in the Via Lactea (VVV) survey\footnote{Detailed information about
  the VVV survey can be found in http://vvvsurvey.org/}, an ESO (European
Southern Observatory) public near-IR variability survey. Its area includes
the Milky Way bulge and an adjacent section of the mid-plane where
star-formation activity is high. This survey will be conducted in the period
$2010-2014$ and will map the whole bulge systematically for multiple epochs.

\begin{figure*}
\includegraphics[bb=-0.1cm .3cm 12cm 6cm,scale=1.2]{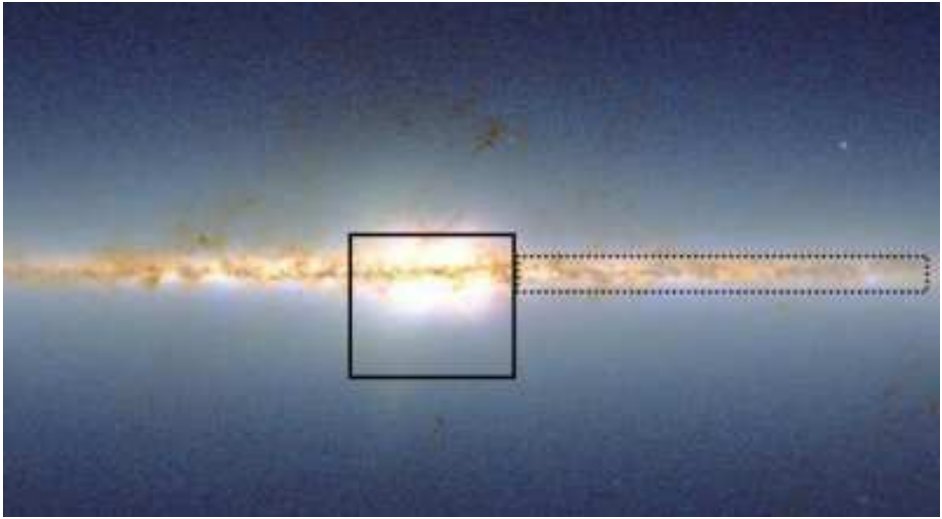}
\caption{2MASS map of the inner Milky Way showing the VVV bulge (solid box, 
$-10 ^{\rm o} < l < +10 ^{\rm o} $ and $ -10 ^{\rm o} < b < +5 ^{\rm o} $) and
plane survey areas (dotted box, $ -65 ^{\rm o} < l < -10 ^{\rm o} $ and $ -2
^{\rm o} < b < +2 ^{\rm o} $).}
\label{area}
\end{figure*}

We plan to cover a 520~deg$^2$ area (Fig.\,\ref{area}) containing
$\sim$$10^9$ point sources. Our survey will give the most complete catalogue
of variable objects in the bulge, with more than $\sim$$10^6$ variables.
Chief among them are the RR~Lyrae, which are accurate primary distance
indicators, and well understood regarding their chemical, pulsational and
evolutionary properties. For the sake of space and coherence we concentrate on
the RR~Lyrae and the star clusters, noting that similar studies can be
done for many of the other populations of variable objects.

Earlier single-epoch near-IR surveys (e.g., COBE, 2MASS, GLIMPSE) have proven
that the Galactic bulge is triaxial and boxy, and contains a bar
\citep{1995ApJ...445..716D,2005AA...439..107L,2005ApJ...630L.149B}. Presently,
the only model we have for the formation of boxy/barred bulges is through
secular evolution of a pre-existing disk. This scenario is believed to be the
dominant channel of formation of bulges in late-type spirals (Sbc), whereas
early-type spiral bulges (S0/Sa) show structural and kinematic evidence for an
early, rapid collapse, which seems to be confirmed by the old age of their
stellar populations \citep[e.g.,][]{2004ARAA..42..603K}.

However, the best-studied spiral bulge, that of the Milky Way, is precisely
the most problematic one to understand in this context. While its surface
brightness shows a barred structure, its stellar population is predominantely
old \citep{2002AJ....124.2054K,2003AA...399..931Z} and has $\alpha$-element
enhancement, characteristic of rapid formation. Nevertheless, the high mean
age of the Bulge still leaves space for a small fraction of young stellar
objects (YSO) which have been found in the inner Bulge
\citep[e.g.,][]{2006AA...453..535S, 2009ApJ...702..178Y}. This is in agreement
with the results of \cite{2006AA...457L...1Z} which indicate that the chemical
composition of the bulge stars is different from that of both thin and
thick-disk stars. Thus, the predictions from the formation of the Milky Way
bulge through secular evolution of the disk seem to be in conflict with some
key properties of its stellar population. However, \cite{2008AA...484L..21M}
recently published results that are in contradiction to
\cite{2006AA...457L...1Z} and show that bulge and disk stars are
indistinguishable in their chemical composition. Given that the near-IR
colours depend strongly on metallicity, the VVV survey will help us to
investigate the metallicity distribution in the survey region. Spectroscopic
data \citep[e.g., future APOGEE;][]{2007AAS...21113208M} will provide
additional $\alpha$-element abundances. 

Our survey of the RR~Lyrae in the Galactic bulge will allow us to map its 3-D
structure \citep[as shown by][]{1995AJ....110.1674C} and will provide key
information on the age of its population, given that RR~Lyrae stars are
tracers of the old population \citep[e.g.,][and references
therein]{2004ApJ...600..409C,2009ApSS.tmp...18C}. This will enable us to
combine the ages of the stellar populations with their spatial distributions. We
note that most single-epoch surveys only provide 2-D maps. With the
present survey, the peak and width of the RR~Lyrae distribution is expected to
be measured with an accuracy of better than $0.01$~mag, which is the required
precision to determine the 3-D structure not only of the bulge, but also of
the Sagittarius dwarf spheroidal galaxy (Sgr dSph) located behind the Milky
Way \citep[e.g.,][]{1996ApJ...458L..17A} and included in our survey. 

At the same time, a comparison between the RR~Lyrae (and type II Cepheids)
in the field and in globular clusters may hold precious information
about the formation of the bulge \citep[e.g.][]{2008MNRAS.386.2115F}. Modern
$\Lambda$CDM cosmology predicts that large galaxies such as the Milky Way
formed by accretion of hundreds of smaller ``protogalactic fragments'', perhaps
not unlike the progenitors of the present-day dwarf spheroidal satellites
\citep[e.g.,][]{2003ApJ...591..499A}. Interestingly, two very massive globular
clusters in the Galactic bulge, NGC~6388 and NGC~6441, have recently been 
suggested to be the remnants of dwarf galaxies that were accreted in
the course of the Galaxy’s history \citep{2002ASPC..265..101R}. These clusters
might prove similar to the cases of M54 (NGC~6715), in the
center of the Sgr dSph, which is currently being cannibalised by the Milky Way
\citep{1995MNRAS.277..781I}, and of $\omega$~Cen (NGC~5139), which has long
been suspected to be the remnant nucleus of a dwarf galaxy \citep[e.g.,][ and
  references therein]{2005AA...439L...5A}. Our proposed search for RR~Lyrae
and type II Cepheids in the Galactic bulge will reveal the presence of
debris related to the accretion events that might have left behind NGC~6441 as
remnant object. The latter is part of our survey.

In order to understand the Milky Way's populations globally, it is necessary
to survey the inner Galactic plane as well. Therefore, we will survey an
adjacent region of the mid-plane and provide a Legacy Database and 3-D atlas
of a large Population I (i.e. young and luminous stars) region. We have
selected the region $-65^{\rm o} < l < -10^{\rm o}$ and $|b| < 2^{\rm o}$ (see
Fig.\,\ref{area}), where star-formation activity is high and for which there
will be complementary optical, mid-IR, and far-IR data from VPHAS$+$, the
Spitzer, GLIMPSE and MIPSGAL surveys, and from the all-sky  AKARI and WISE
survey. The addition of this region will also permit us to discriminate
between various models of the inner Galactic structure which, besides the
triaxial bulge, contain a long bar and a ring \citep[e.g.,][ and references
therein]{2007AJ....133..154L} or not \citep[e.g.,][ and references
therein]{2004ASPC..317..289M}. Indeed, the selected region includes the
putative negative-longitude tip of the long bar (at $l\approx -14^\circ $,
$|b|<1^\circ $), which has not yet been observed.

The large survey area will allow several remaining astrophysical problems to
be addressed. For example, the effect of the environment on star formation and
in particular the initial mass function (IMF) at low masses is presently
poorly known. This issue will be addressed statistically by observing hundreds
of star-forming regions and cross-correlating the shapes of their luminosity
functions with cluster density, the presence of high-mass stars, and
galactocentric distance. For comparison, VVV survey will reach $1~mag$ deeper
than UKIDSS Galactic Plane Survey (GPS), which overlaps with VVV in the region
of $-2^\circ<l<+10^\circ$, $|b|<2^\circ$. Other important parameters, such as
velocity dispersion and metallicity, will be determined by spectroscopic
follow-up observations. In addition, the luminosity function of the clusters
themselves will be measured, for both star-forming clusters and more evolved
open clusters.  

These issues cannot be addressed with optical surveys, owing to the high
extinction in the plane. The Spitzer data will be invaluable for detecting the
most obscured high-mass protostars within star-forming regions. A near-IR
survey will be more sensitive to all but the reddest objects, and the superior
spatial resolution in these wavebands will be essential for resolving distant
clusters and the crowded field populations.

\section{Technical description}

\subsection{Telescope and instrument design}

The Visible and Infrared Survey Telescope for Astronomy (VISTA) is a 4m-class
``wide-field'' telescope located at ESO's Cerro Paranal Observatory in Chile,
designed to conduct large-scale surveys of the southern sky at near-IR
wavelengths (0.9 to 2.5~$\mu$m). The telescope has an altitude-azimuth mount,
and quasi Ritchey-Chr\'etien optics. An $f/1$ primary mirror was designed
together with Cassegrain-focus instrumentation to offer the best solution to
the difficult problem of combining a wide-field with good image quality, and
results in a physically large focal plane with an $f/3.25$ focus
\citep{2006SPIE.6267E...7M}. VISTA's active optics uses two low-order
curvature sensors, which operate concurrently with science exposures, and a
high-order curvature sensor.

\begin{figure}[ht]
\includegraphics[bb=-4cm 4.5cm 12cm 21.3cm,scale=0.60]{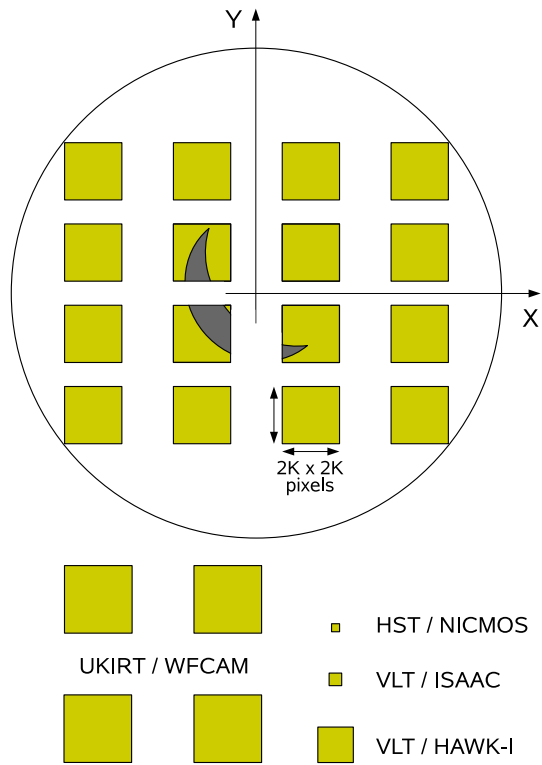}
\caption{Diagram showing the array of the sixteen detectors on the VISTA
camera and the axis orientation used to shift the camera in each exposure
to obtain the tiles. For comparison we show the crescent Moon over the VISTA
camera and the fields of view of UKIRT/WFCAM, HST/NICMOS, VLT/ISAAC,
and VLT/HAWK-I.}  
\label{camera}
\end{figure}

The telescope is equipped with a near-IR camera containing 67 million pixels
(an array of $16\times2048\times2048$ Raytheon VIRGO IR detectors) of mean
size $0\farcs34$ and available broad-band filters at $Z\,Y\,J\,H\,K_{\rm s}$
and a narrow-band filter at 1.18 $\mu$m. Given VISTA's nominal pixel size, the
diameter of the field of view is 1.65~deg. The point-spread function (PSF)
of the telescope+camera system (including pixels) is designed to have a full
width at half maximum (FWHM) of $0\farcs51$, not including the contribution of
atmospheric turbulence. Seeing, and other weather-related statistics for Cerro
Paranal, are given at ESO's ``Astroclimatology of Paranal'' web
pages\footnote{http://www.eso.org/gen-fac/pubs/astclim/paranal/}. The VISTA
site is expected to have similar conditions, which are well suited to our
survey requirements. 

The 16 detectors in the camera are not buttable and are arranged as shown
in Figure\,\ref{camera}. Each individual exposure
produces a sparsely sampled image of the sky known as a ``pawprint'', covering
an area of 0.599~deg$^2$.

To `fill in' the gaps between the detectors to produce a single filled
``tile'' with reasonably uniform sky coverage, the minimum number of pointed
observations (with fixed offsets) required is six (three offsets in $Y$ and
two offsets in $X$). After six steps an area of $1.501$~deg$^2$ on the sky,
corresponding to one tile, is (almost) uniformly covered.

\subsection{Data reduction}

We will use the enhanced VISTA Data Flow System\footnote{The VISTA Data Flow
System (VDFS) is a collaboration between the UK Wide Field Astronomy Unit at
Edinburgh (WFAU) and Cambridge Astronomy Survey Unit (CASU), coordinated by
the VISTA PI and funded for VISTA by the Science and Technology Facilities
Council.}
\citep[VDFS;][]{2004SPIE.5493..401E,2004SPIE.5493..411I,2004SPIE.5493..423H}.
It includes all basic data reduction steps:

(i) removing instrumental signature (bias and dark frames, twilight, and
dome flatfields, linearity, bad pixel maps, cross-talk, gain calibrations),
merging pawprints into tiles and calibrating photometrically and
astrometrically;

(ii) extracting source catalogues on a tile-by-tile basis; 

(iii) constructing survey-level products -- stacked pixel mosaics, difference
images, and merged catalogues; 

(iv) providing the team with both data access and methods for querying and
analyzing the data; and 

(v) producing virtual observatory (VO)-compliant data products for delivery to
the ESO archive. 

\begin{figure}
\includegraphics[bb=-1.8cm 7.5cm 12cm 24.5cm,scale=0.48]{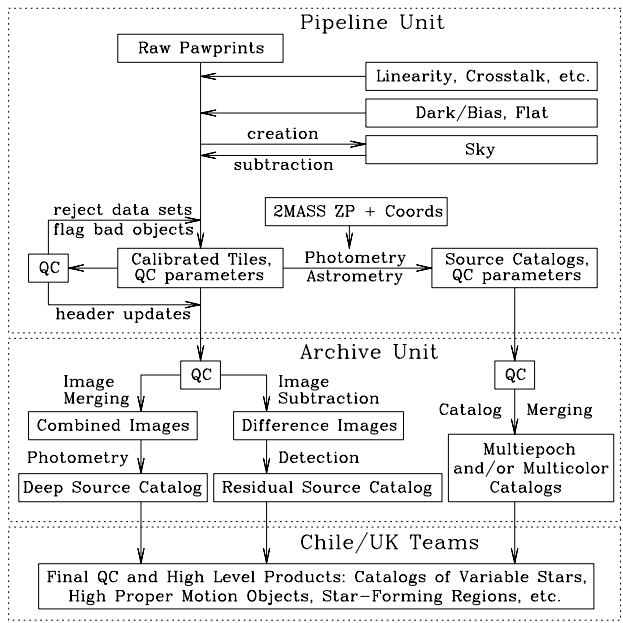}
 \caption{Flow chart of the VVV data processing (QC: Quality Control; ZP:
   Zeropoint).}
\label{flow}
\end{figure}

Figure\,\ref{flow} shows a flow chart of the data processing. The pipeline
products are: astrometrically corrected and photometrically calibrated tiles
in each filter used, confidence maps, and homogeneous object catalogues
\citep{2009MNRAS.399.1730C}. The pipeline records the processing history and
calibration information of each file, including calibration files and quality
control parameters. The Cambridge Astronomy Survey Unit (CASU) component of
the VDFS will be responsible for the basic pipeline processing and the first
calibration, all done on a daily basis. 

\subsection{Combination/image subtraction (archive)}

The ``second''-order data processing requires access to larger sets of
data to produce survey products.  It is carried out by the Wide Field Astronomy
Unit's (WFAU) VISTA Science Archive (VSA) in Edinburgh. The Science Archive
contains only calibrated data and catalogues, and no raw data.
The Science Archive is responsible for: 

(i) image stacking to produce combined and differenced tiles and source
merging;

(ii) quality control: assessment of the data quality and filtering of the
data that do not meet the established criteria for photometric and astrometric
accuracy; 

(iii) light-curve extraction: this will be done by implementing an
image-subtrac- tion algorithm \citep{1998ApJ...503..325A,2000AAS..144..363A},
which will allow us to create the catalogue of variable sources. This method
provides excellent results for crowded fields in which the traditional
aperture or PSF-fitting photometry fails \citep[e.g.,
][]{2004AA...424.1101K,2005AcA....55..261P}.

\subsection{Photometric calibration}

During the first period we will carry out the external calibrations and
transformations to the standard system using 2MASS and UKIDSS for
bootstrapping\footnote{The filter transmission curves for each instrument can be
  found at http://www.vista.ac.uk/index.html (VISTA),
  http://web.ipac.caltech.edu/staff/waw/2mass/opt\_cal/in- dex.html (2MASS)
  and http://www.ukidss.org/technical/instrument/filters.html (UKIDSS).}.

The calibration scheme for a given filter is as follow:
\begin{equation}
m_{\rm cal} = m_{\rm inst} + ZP -k(X-1) = m_{\rm std} + clr_{\rm std}
\end{equation}
where $m_{\rm cal}$ is the calibrated magnitude, $m_{\rm inst}$ the measured
instrumental magnitude, $ZP$ the zeropoint, $k$ the extinction coefficient, and
$X$ the airmass of the object. On the right-hand side of this equation,
$m_{\rm std}$ and $clr_{\rm std}$ are the corresponding standard magnitude and
colour.

Calibration and quality control is done using 2MASS stars in the frames
themselves, applying colour equations to convert 2MASS photometry to the VISTA
photometric system \citep{2006AJ....131.1163S, 2009MNRAS.394..675H}. 

There are thousands of unsaturated 2MASS stars in $J\,H\,K_{\rm s}$ with
photometric errors $<$0.1~mag in every VISTA tile field. A large fraction of
these can be sufficiently isolated even in the crowded fields.

To calibrate the $Y-$ and $Z-$band data (both filters are not available from
2MASS) we will use observations of the standard VISTA calibration fields as
required by the ESO Public Survey Panel. Details will be published in a
forthcoming paper describing the science verification.

The internal gain correction applied through flat-fielding will place the
detectors on a common zero-point system. After deriving this $ZP$ in each tile,
a double check using the overlap regions will be made to estimate the internal
photometric accuracy.

\begin{figure}[ht!]
\includegraphics[bb=-1.4cm 0.5cm 16cm 21.3cm,scale=0.52]{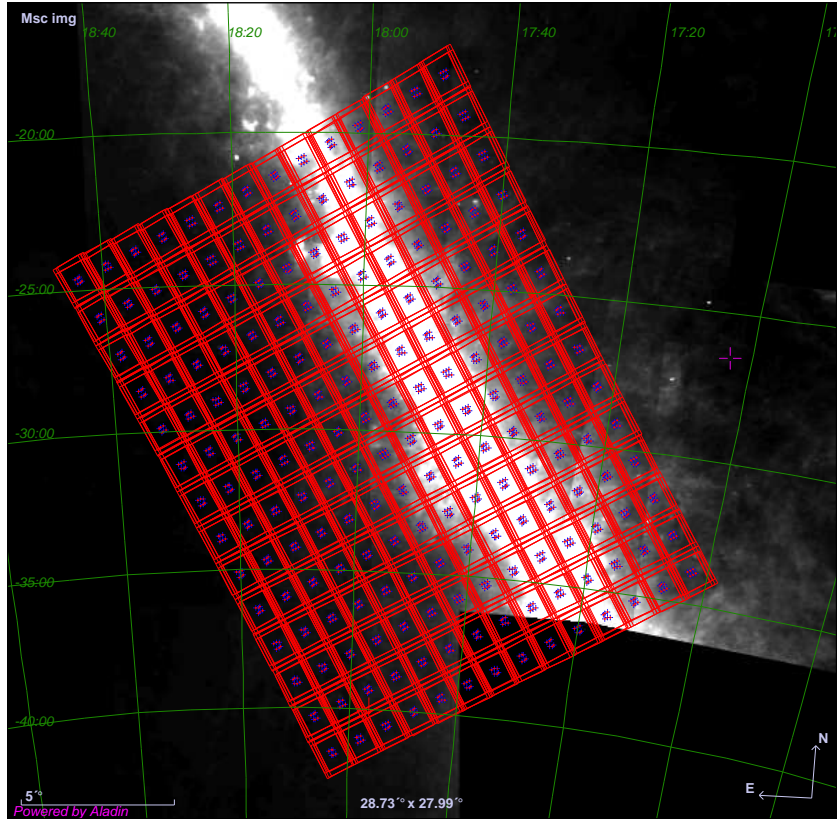}
\caption{Coverage of the Galactic center region overlaid on a mid-IR
map. Boxes mark the tiles needed to cover the bulge area (see
Fig.\,\ref{area}), whereas the crosses mark the box centers.}
\label{tiles}
\end{figure}

\section{Observing strategy}

The VISTA tile field of view is $1.501$~deg$^2$, hence 196 tiles are needed to
map the bulge area and 152 tiles for the disk\footnote{The tiles' spacing and
orientation were calculated with the Survey Area Definition Tool (SADT)
software to maximize the efficiency of the sky coverage. See
http://www.vista.ac.uk/observing/sadt/}. Adding some $X$ and $Y$ overlap
between tiles for a smooth match, the area of our unit tile  covered twice is
$1.458$~deg$^2$. Figure\,\ref{tiles} provides a schematic representation of
the tiling scheme for the Galactic center region.

The variability study in the bulge will be carried out in the $K_{\rm s}$ band
down to $\sim$18~mag (signal-to-noise $\approx3$). The total exposure time for
a VISTA tile field is 162~s.  Our strategy yields about 30~deg$^2$ per hour,
or 300~deg$^2$ per night. The combined epochs will reach $K_{\rm s} = 20$~mag,
which is three magnitudes fainter than the unreddened bulge main-sequence
turn-off (MS turn-off), although the densest fields will be
confusion-limited. However, applying both PSF fitting and image subtraction,
we will recover the light curves of most objects down to $K_{\rm s} = 18$~mag,
even in moderately crowded fields. This is more than 3~mag fainter than the
unreddened known RR~Lyrae in the Galactic bulge. We expect to find RR~Lyrae
even in fields with $A_{V} = 10$~mag.

Table\,\ref{mags} lists some reference $K_{\rm s}$-band magnitudes at the
distance of the bulge for a range of extinction and reddening values. These
typical magnitudes were obtained from \cite{1995AJ....110.1674C},
\cite{1996ApJ...458L..17A}, \cite{1998ApJ...492..190A}, and
\cite{2003AA...399..931Z}. As a reference point, for Baade's window $E(B-V) =
0.5$~mag, so that $A_{V} = 1.5$, $A_{J} = 0.4$, and $A_{K} = 0.2$~mag
\citep{1985ApJ...288..618R}. This table shows that the tip of the bulge
red-giant branch (RGB) will saturate ($K_{\rm s}<9.5$), but for the RGB clump
giants, and even for the tip of the RGB of the Sgr dSph galaxy, the VVV survey
will be able to see giants throughout the bulge, even in the most obscured
regions. The bulge RR~Lyrae and the Sgr dSph galaxy red-clump giants will also
be detected, even for the regions with the highest extinction ($A_{V} >
30$~mag) at low Galactic latitudes. Finally, the RR~Lyrae of the Sgr dSph
galaxy and the bulge MS turn-off stars will be detected only in the regions
with low absorption ($A_{V} < 10$~mag) at higher latitudes. We are aware that
the use of an `universal' extinction law $A_V$=$3.1$~$E(B-V)$ is problematic
in the inner region of the bulge. According to \cite{2006ApJ...638..839N} and
\cite{2009MNRAS.394.2247G} a single extinction law is not consistent with the
observations of the Galactic center along different lines of sight.  

\begin{sidewaystable}
\caption{$K_{\rm s}$-band magnitudes at the distance of the bulge. The
  absorption values are based on the standard extinction law as derived by
  \cite{1985ApJ...288..618R}.}
\vspace{12pt}
\begin{tabular}{l c c c c c c}
\hline
           & {\small $E(B-V)$=0}  & {\small $E(B-V)$=0.5} & {\small $E(B-V)$=1.5} & {\small $E(B-V)$=3.2} & {\small $E(B-V)$=4.8} & {\small $E(B-V)$=8.4}\\
           & $A_{V} = 0$ & $A_{V} = 1.5$ & $A_{V} = 5.0$ & $A_{V} = 10.0$ & $A_{V} = 15.0$ & $A_{V} = 26.3$\\
           & $A_{J} = 0$ & $A_{J} = 0.4$ & $A_{J} = 1.4$ & $A_{J} =  2.8$ & $A_{J} = 4.2$  & $A_{J} = 7.4$\\
{\small Population} & $A_{K} = 0$ & $A_{K} = 0.2$ & $A_{K} = 0.6$ & $A_{K} =  1.1$&$A_{K} = 1.7$  & $A_{K} = 3.0$\\
\hline
\hline
{\small Bulge RGB tip}  &   $K_{\rm s}=8.0$* & $K_{\rm s}=8.2$*  & $K_{\rm s}=8.6$*  & $K_{\rm s}=9.1$* &$K_{\rm s}=9.7\,\,\,$ & $K_{\rm s}=11.0$\\
{\small Sgr dSph RGB tip}    &   $K_{\rm s}=10.5$ & $K_{\rm s}=10.7$  & $K_{\rm s}=11.1$  & $K_{\rm s}=11.6$ & $K_{\rm s}=12.2 $  &$K_{\rm s}=13.5$\\
{\small Bulge RGB clump} &   $K_{\rm s}=12.9$ & $K_{\rm s}=13.1$  & $K_{\rm s}=13.5$  & $K_{\rm s}=14.0$ & $K_{\rm s}=14.6 $  &$K_{\rm s}=15.9$\\
{\small Bulge RR Lyrae} &   $K_{\rm s}=14.3$ & $K_{\rm s}=14.5$  & $K_{\rm s}=14.9$  & $K_{\rm s}=15.4$ & $K_{\rm s}=16.0 $  &$K_{\rm s}=17.3$\\
{\small Sgr dSph RGB clump}  &   $K_{\rm s}=15.4$ & $K_{\rm s}=15.6$  & $K_{\rm s}=16.0$  & $K_{\rm s}=16.5$ & $K_{\rm s}=17.1 $  &$K_{\rm s}=18.4$**\\
{\small Sgr dSph RR Lyrae}   &   $K_{\rm s}=16.8$ & $K_{\rm s}=17.0$  & $K_{\rm s}=17.4$  & $K_{\rm s}=17.9$ & $\,\,\,K_{\rm s}=18.5$**  &$K_{\rm s}=19.8$**\\
{\small Bulge MS turn-off}    &   $K_{\rm s}=17.0$ & $K_{\rm s}=17.2$  & $K_{\rm s}=17.6$  & $K_{\rm s}=18.1$ & $\,\,\,K_{\rm s}=18.7$**  &$K_{\rm s}=20.0$**\\
\hline
\label{mags}
\vspace{-0.5cm}
{\small ~*~saturated}
\vspace{12pt}
\\
{\small ** beyond detection}
\end{tabular}
\end{sidewaystable}

For the plane survey, the $K_{\rm s}$-band observations require a total time
of 80~s on target, and an elapsed time of 366~s per tile. 

Bright point sources with $K_{\rm s} < 9.5$~mag will be saturated in the
individual images. This will therefore include most unreddened bulge Mira
variables, but Miras in the Sgr dSph galaxy can be monitored, as well as Miras
located in regions with very high extinction (e.g., next to the Galactic
center). The Mira population in the Galactic center has been studied by
\cite{2009MNRAS.399.1709M}. In addition, bright-star saturation may be an
issue, but we estimate that even in the worst cases only a small portion of
the field will be rendered useless. Hence we do not expect that the
saturation of the brightest stars to effect our conclusions about
three-dimensional structure of the inner Milky Way. For example, in the
optical microlensing surveys where CCD bleeding is comparatively worse, less
than 5\% of the most crowded bulge fields are lost.

\begin{figure}[!ht]
\includegraphics[bb=.2cm 6.2cm 16cm 18.3cm, scale=0.60]{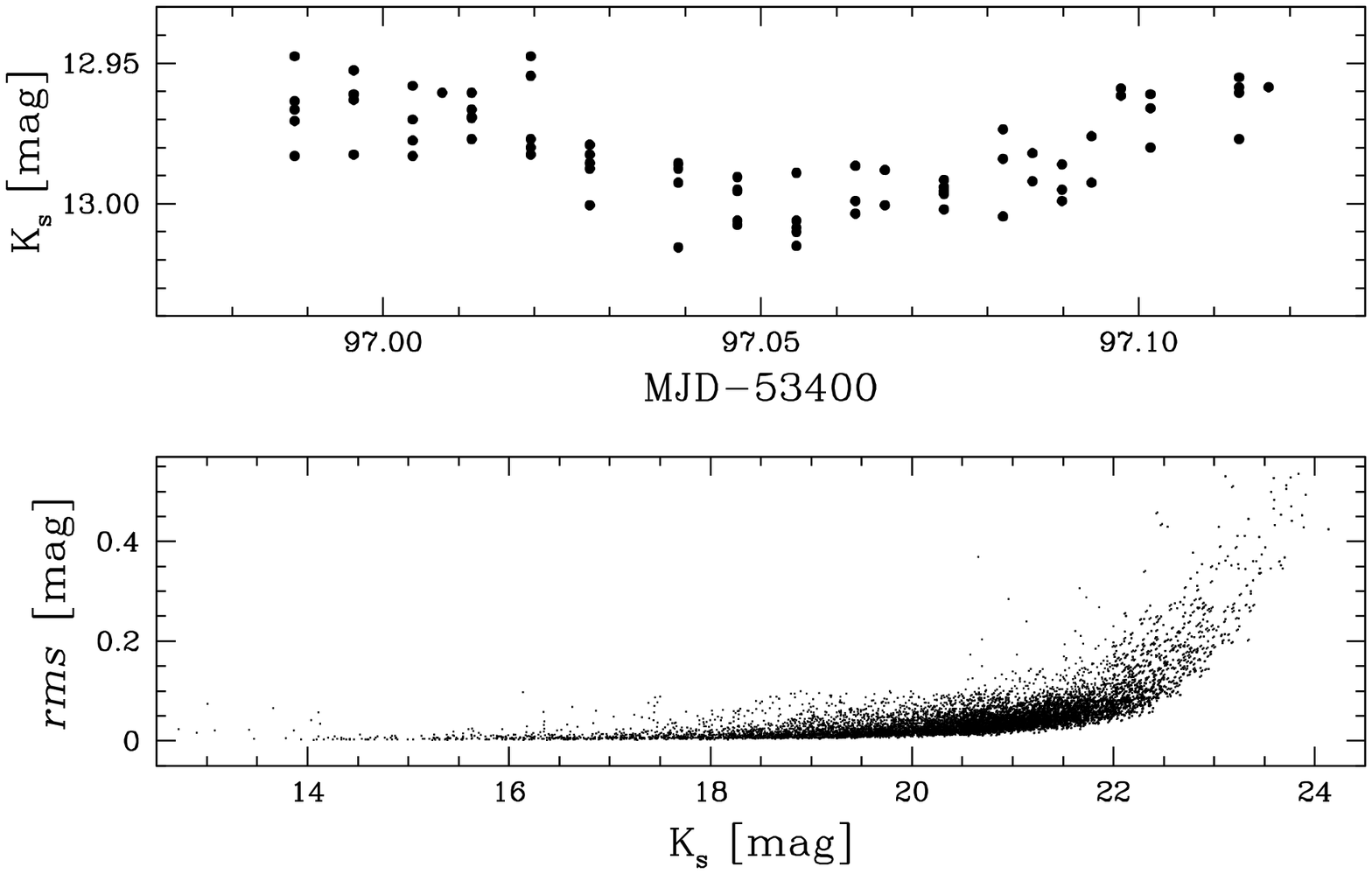}
\caption{Top panel: Light curve of the planetary transit OGLE-TR-113
measured with SOFI at the 3.5-m ESO NTT telescope in the $K_{\rm s}$ band.
This star is located in the Galactic plane in the field of Carina, a moderately
crowded region. Bottom panel: Precision of the relative photometry obtained
with SOFI as a function of magnitude. The magnitudes and $rms$ are calculated
in two iterations, removing the $> 10\sigma$ outliers.}
\label{tr113}
\end{figure}

To illustrate the precision of crowded-field IR photometry we include
Fig.\,\ref{tr113}, showing photometry of the planetary transit OGLE-TR-113
obtained with NTT/ SOFI (top panel), and photometric accuracy of those
observations as a function of magnitude (bottom panel). In order to evaluate
the amplitude threshold for our detections, we have carried out Monte Carlo
simulations using the RR Lyrae light curve templates from
\citet{1996PASP..108..877J} and \citet{2005AJ....129.2714D}. As a result, we
find that, at a typical magnitude of $K_s \approx 15-16$, and taking into
account the expected photometric errors, we should be able to detect RR Lyrae
stars with amplitudes down to $A_K = 0.05-0.07$~mag using 80 datapoints from
the first three years of VVV operation, and further down to $A_K =
0.03-0.05$~mag if the dataset is extended to cover 180 phase points over a
time frame of 5 years.

\begin{figure}[ht!]
\includegraphics[bb=2.0cm 11.5cm 10cm 18cm,scale=1.00]{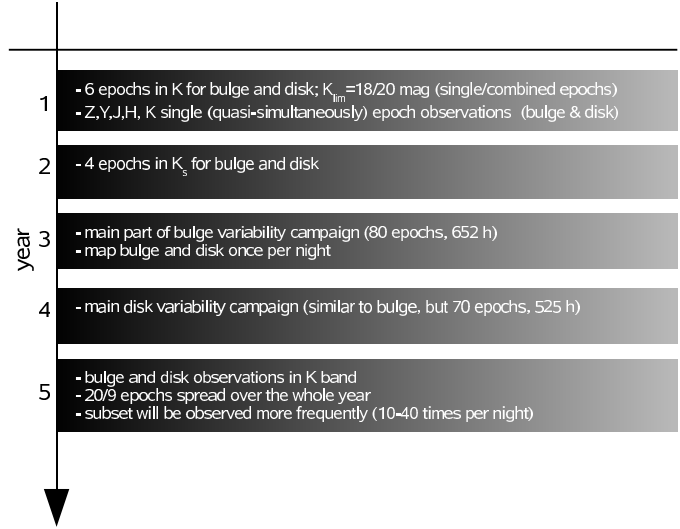}
\caption{Schematic schedule of the VVV survey strategy.}
\label{schedule}
\end{figure}

The total estimated time per observing period is given in Table\,\ref{obs},
which also shows the requirements for Moon, seeing, and transparency
conditions. The times include overheads (both for readout and for changing to
a new tile) and time possibly spent on standard stars for the $Z$ and $Y$
observations (not used in the 2MASS and UKIDSS survey).  This strategy allows
us to provide various data to the community, enabling follow-up throughout the
survey. The full survey will require a total of 192 nights of observations
over 5 years. A schematic schedule of the survey is shown in
Fig. \ref{schedule}.

During the first year, the whole bulge area will be observed in the
$K_{\rm s}$ band for 6 consecutive epochs, for a total of 65~hours.
A further 86~hours will be devoted to complete imaging of each bulge tile in
$Z\,Y\,J\,H$. This will provide reliable near-simultaneous
fluxes and colours for each tile area. 

The same strategy will be applied to the 152 tiles covering the disk area for
the single-epoch and the quasi-simultaneous multi-colour disk survey. The
total time spent on the disk for the first year is thus 141~hours. Added to
the 151~hours for the bulge, we will thus spend 292~hours in total on the
survey during the first year. The multi-color observations, in combination
with datasets from  UKIDSS GPS (near IR), VST/VPHAS+ (optical) and GLIMPSE and
GLIMPSE-II (mid-IR), will be used to build improved extinction maps for the
survey region. Note that the individual, single-epoch observation blocks (OBs)
in $K_{\rm s}$ all have the same limiting depth (under the same conditions),
whether forming part of the VVV's bulge or disk components. 

\begin{table*}[ht!]
\begin{center}
\caption{Estimated observing time for the full survey.}
\vspace{12pt}
\begin{tabular}{c c c c c c l}
\hline
Year & Time  & RA range & Moon & Seeing & Transparency & ~~~~~~Number of \\
     &  [h]  &  [hh:mm]     &      & [~$\arcsec$~] &  & ~~~~K-band epochs\\
\hline
\hline
year 1 & ~~292 & 12:00--19:00  & any & 0.8 & clear & 6 (bulge and disk)\\
year 2 & ~~292 & 12:00--19:00  & any & 0.8 & clear & 4 (bulge and disk)\\
year 3 & ~~652 & 12:00--19:00  & any & any & thin  & 80 (bulge only)\\
year 4 & ~~525 & 12:00--19:00  & any & any & thin  & 70 (disk only)\\
year 5 & ~~168 & 12:00--19:00  & any & 0.8 & clear & 20 (bulge) + 9 (disk)\\
\hline
Total  & 1929 &              &     &     &      & \\
\hline
\label{obs}
\end{tabular}
\end{center}
\end{table*}

Being fully aware of the confusion and background limits, the observing plan
would cycle alternately through fields of varying density for optimal sky
subtraction. The filter order in the OBs will be optimized to minimize
overheads.

During the second year, we will acquire another 20 epochs in $K_{\rm s}$ for
the whole bulge (217~hours) and 10 for the plane (75~hours), for a total of
292~hours. These additional epochs will improve our ability to detect variable
sources (but will not permit us to conclusively establish the variability
phases, at least for many of the RR~Lyrae stars). These data will also allow
the creation of deeper master maps in $K_{\rm s}$, to fine tune the strategy
for the main campaign of the following year.

During the third year, the main bulge-variability campaign of 60 epochs will
be carried out over 652~hours. According to the ESO Public Survey
requirements, only 40 nights can be consecutive, while the others will be
spread over the bulge season. We will use the $K_{\rm s}$ band to map the whole
bulge and inner plane. A subset of the fields can be observed more frequently
(4$ - 8$ times per night). This strategy allows us to partially remove
aliasing and to improve the periods, while being more sensitive to shorter
timescale variables and microlensing events.

During the fourth year, the main Galactic-plane variability campaign will be
carried out over 70 epochs using 525~hours in $K_{\rm s}$, following a similar
strategy as for the bulge in the previous year. According to the ESO
requirements, only part of these observations can be carried out over 27
consecutive nights.

Finally, during the fifth year, we will acquire 12 more epochs for the bulge
and five more for the disk, with observations spread over the season, taking a
total of 130 and 38 hours, respectively. This allows measurements of
longer-timescale variables, and the search for high-proper-motion objects. A
subset of the fields can be observed much more frequently (10--40 times per
night). This strategy enables detection of short-period variables and
planetary transits.

\section{Scientific goals}

The major VVV survey products will be a high-resolution $Z\,Y\,J\,H\,K_{\rm s}$
colour atlas of the bulge and plane regions, and a catalogue of variable point
sources, including positions, mean magnitudes, and amplitudes. We expect to
detect more than $10^6$ variable objects. However, the total number may even 
reach $10^7$ if one takes into account results from recent deep variable
searches in optical bands in selected Galactic-plane regions
\citep{2008AJ....135..649W,2009AA...503..651P}. Our database will be public,
a significant treasure for the whole community to exploit for a variety of
scientific programmes.  
\\
\\
The main scientific goals of the VVV survey are:
\\
\\
1. {\it To find RR~Lyrae in the bulge}, which will allow us to determine
periods and amplitudes, and measure accurate mean $K_{\rm s}$ magnitudes. We
will construct the Bailey diagram (luminosity amplitude {\em vs.} period), and
interpret the results of the variability analysis in terms of stellar
pulsation and evolution models, similarly to what is currently done using
(primarily) the visual bandpasses \citep[e.g.,][]{2000ApJ...532L.129B,
2007AA...476..779B, 2003ApJ...596..299M, 2005AJ....129..267C,
2007AA...474..557M}.  As pointed out by \citet{2000ApJ...532L.129B}, one
common limitation in the comparison between hydrodynamical pulsation models
and the observations is the lack of proper sampling in the available near-IR
light curves~-- a situation which will be dramatically improved with the
advent of the well-sampled $K_{\rm s}$-band light curves from the VVV
survey. Naturally, while RR Lyrae stars are the main focus of this project,
similar such studies will be possible for many other types of variable stars,
provided they are detected in significant numbers in our studied fields~--
including, for instance, classical Cepheids \citep{2007AA...476..863F}, type
II Cepheids \citep{2007AA...471..893D}, anomalous Cepheids
\citep{2006AA...460..155F}, and Miras and semiregular variables
\citep{2006AA...460..539K, 2006AJ....131..612S}, among others~-- with a
significant impact also upon their use as distance indicators.

\begin{figure}[ht]
\includegraphics[bb=-3.0cm 0.8cm 10cm 10.5cm,scale=0.78]{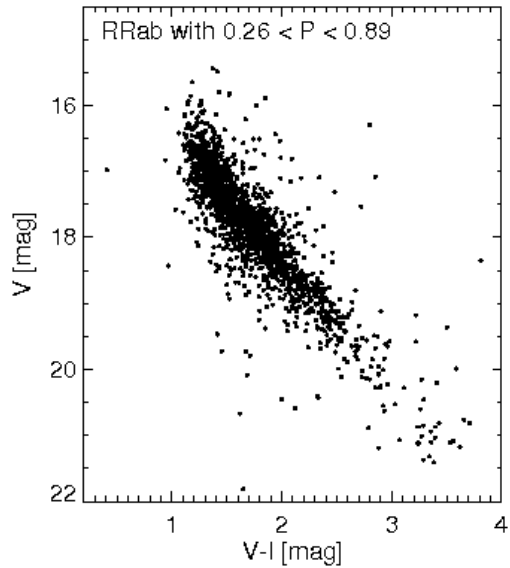}
\caption{Optical colour-magnitude diagram for fundamental mode RR Lyrae stars
  in the Galactic bulge from the Optical Gravitational Lensing Experiment
  \citep[][]{2006ApJ...651..197C}}
\label{alcock}
\end{figure}

The pulsation properties of bulge variables will be compared with those of
similar variables in the halo and nearby dwarf galaxies
\citep[e.g.,][]{2009ApSS.tmp...18C}. The distances measured and RR Lyrae
counts can be compared with the red-clump giants, which are excellent tracers
of the inner bar \citep{1994ApJ...429L..73S}. This would define the geometry
of the inner bar and of additional structures \citep[such as a potential
second bar;][]{2005ApJ...621L.105N}, and explore the radial dependence of the
density \citep[e.g.,][]{1999ASPC..165..284M}, or trends with Galactic
latitude-longitude, to finally unveil the structure of the bulge. The
microlensing surveys that we have been involved in (OGLE and MACHO) have
discovered a small fraction of the bulge RR~Lyrae stars \citep[e.g.,
Fig.\,\ref{alcock}, see][]{1998ApJ...492..190A,2002AcA....52..129W}. Our
survey will increase the amount of data on bulge RR~Lyrae significantly.

\begin{figure*}[!ht]
\includegraphics[bb=0.6cm 0.5cm 21cm 8.8cm,scale=0.65]{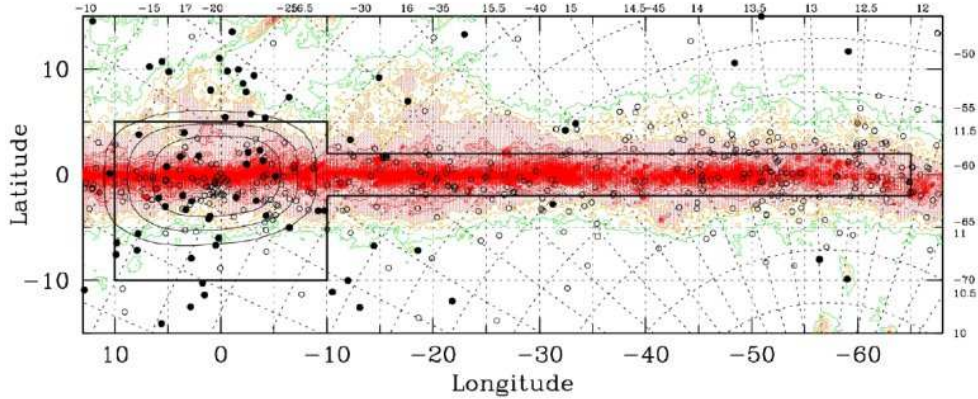}
\caption{Map of the globular and open cluster positions (full and empty circles,
respectively) towards the Milky Way bulge and plane. Included in the VVV area
are the 33 globular clusters listed in Table~3 \citep{1996AJ....112.1487H},
and 355 open clusters
\citep{2003AA...404..223B,2003AA...400..533D,2006AA...446..949D}. The bulge
contours are indicated.}
\label{clusters}
\end{figure*}

We will search for RR~Lyrae and type II Cepheids in the Galactic bulge to
reveal the presence of any debris related to accretion events that might have
left behind the present-day globular cluster NGC~6441.  This globular cluster
is well known to contain an anomalous RR~Lyrae population, with periods that
are much longer than those of known field RR~Lyrae stars of similarly high
metallicity \citep[e.g.,][]{2000ApJ...530L..41P,2003AJ....126.1381P}. In
particular, the presence of the unusually long-period ($P>0.45$~d) RRc (first
overtone) variables, which have so far not been found in the general field but
are present in large numbers in this globular cluster
\citep[e.g.,][]{2004ASPC..310..113C}, should provide the “smoking gun” for the
presence of NGC~6441-related debris in the general bulge field.  In a similar
vein, long-period RRab stars (fundamental-mode pulsators) occupying the
appropriate position in the period-amplitude diagram should also provide us
with a strong indication of prior membership of such a protogalactic
fragment. 
\\
\\
2. {\it To identify variable stars belonging to known star clusters.} There
are 33 globular clusters and 355 open clusters located in the VVV area
(Fig.\,\ref{clusters}), which may contain RR~Lyrae, type I and II Cepheids,
semiregular variables, and eclipsing binaries, among other types of
variables. Distances, reddening values, metallicities, and horizontal-branch
(HB) types will be obtained for these clusters from a homogeneous dataset
\citep[e.g.,][]{2006MmSAI..77..202C,2003AA...399..931Z}. In some favourable
cases, ages can be derived. Table\,\ref{globular} lists the globular clusters
that will be covered, giving positions in equatorial and Galactic coordinates,
and distances from the Sun. The asterisks in the last column indicate that
more than one third of these clusters have uncertain distances. We will
improve the distances for these globulars, and confirm the previous estimates
for the rest of the open and globular cluster sample. 

\vskip 1.5em

\begin{longtable}{l c c c c c}
\caption{The 33 known globular clusters within the VVV area.}\\
\hline
              & RA (J2000) & DEC (J2000) &  $l$   &  $b$     & D      \\
Cluster ID    & [hh:mm:ss] & [dd:mm:ss]  & [deg]  & [deg]    & [kpc] \\
\hline
\hline
Terzan 2  & 17:27:33.1 & --30:48:08 & 356.32 &  2.30 &  8.7\,\,\,\, \\
Terzan 4  & 17:30:39.0 & --31:35:44 & 356.02 &  1.31 &  9.1\,*      \\
HP 1      & 17:31:05.2 & --29:58:54 & 357.42 &  2.12 & 14.1\,*      \\
Liller 1  & 17:33:24.5 & --33:23:20 & 354.84 & --0.16 &  9.6\,*      \\
NGC 6380  & 17:34:28.0 & --39:04:09 & 350.18 & --3.42 & 10.7\,\,\,\, \\
Terzan 1  & 17:35:47.2 & --30:28:54 & 357.56 &  0.99 &  5.6\,\,\,\, \\
Ton 2     & 17:36:10.5 & --38:33:12 & 350.80 & --3.42 &  8.1\,*      \\
NGC 6401  & 17:38:36.6 & --23:54:34 & 3.45   &  3.98 & 10.5\,\,\,\, \\
Pal 6     & 17:43:42.2 & --26:13:21 & 2.09   &  1.78 &  5.9\,\,\,\, \\
Djorg 1   & 17:47:28.3 & --33:03:56 & 356.67 & --2.48 & 12.0\,*      \\
Terzan 5  & 17:48:04.9 & --24:46:45 & 3.84   &  1.69 & 10.3\,*      \\
NGC 6440  & 17:48:52.7 & --20:21:37 & 7.73   &  3.80 &  8.4\,\,\,\, \\
NGC 6441  & 17:50:12.9 & --37:03:05 & 353.53 & --5.01 & 11.7\,\,\,\, \\
Terzan 6  & 17:50:46.4 & --31:16:31 & 358.57 & --2.16 &  9.5\,*      \\
NGC 6453  & 17:50:51.7 & --34:35:57 & 355.72 & --3.87 &  9.6\,\,\,\, \\
UKS 1     & 17:54:27.2 & --24:08:43 & 5.12   &  0.76 &  8.3\,*      \\
Terzan 9  & 18:01:38.8 & --26:50:23 & 3.60   & --1.99 &  6.5\,*      \\
Djorg 2   & 18:01:49.1 & --27:49:33 & 2.76   & --2.51 &  6.7\,*      \\
Terzan 10 & 18:02:57.4 & --26:04:00 & 4.42   & --1.86 &  5.7\,*      \\
NGC 6522  & 18:03:34.1 & --30:02:02 & 1.02   & --3.93 &  7.8\,\,\,\, \\
NGC 6528  & 18:04:49.6 & --30:03:21 & 1.14   & --4.17 &  7.9\,\,\,\, \\
NGC 6540  & 18:06:08.6 & --27:45:55 & 3.29   & --3.31 &  3.7\,\,\,\, \\
NGC 6544  & 18:07:20.6 & --24:59:51 & 5.84   & --2.20 &  2.7\,\,\,\, \\
NGC 6553  & 18:09:17.6 & --25:54:31 & 5.25   & --3.03 &  6.0\,\,\,\, \\
HJK2000-GC02 & 18:09:36.5 & --20:46:44 & 9.78   & --0.62 &  4.0\,*      \\
NGC 6558  & 18:10:17.6 & --31:45:50 & 0.20   & --6.02 &  7.4\,\,\,\, \\
Terzan 12 & 18:12:15.8 & --22:44:31 & 8.36   & --2.10 &  4.8\,*      \\
NGC 6569  & 18:13:38.8 & --31:49:37 & 0.48   & --6.68 & 10.7\,\,\,\, \\
NGC 6624  & 18:23:40.5 & --30:21:40 & 2.79   & --7.91 &  7.9\,\,\,\, \\
NGC 6626  & 18:24:32.9 & --24:52:12 & 7.80   & --5.58 &  5.6\,\,\,\, \\
NGC 6638  & 18:30:56.1 & --25:29:51 & 7.90   & --7.15 &  9.6\,\,\,\, \\
NGC 6642  & 18:31:54.1 & --23:28:31 & 9.81   & --6.44 &  8.4\,\,\,\, \\
NGC 6656  & 18:36:24.2 & --23:54:12 & 9.89   & --7.55 &  3.2\,\,\,\, \\
\hline
* uncertain distances
\label{globular}
\end{longtable}

\vskip 1.5em

For some of the Cepheids which we expect to find in open clusters, we will
derive improved distances by applying the IR surface-brightness
technique \citep{2005ApJ...627..224G,1997AA...320..799F} on these variables. 
\\
\\
3. {\it To find eclipsing binaries in large numbers.} We expect to detect more
than $5\times10^{5}$ binaries, an unprecedented database that will allow us to
determine periods, amplitudes, mean magnitudes, study stellar properties, and
also select extrasolar planetary-transit candidates. In particular,
YY~Gem-like systems can be identified to constrain the lower main-sequence
parameters \citep[e.g.,][]{2002ApJ...567.1140T}, and selected transit fields
can be followed frequently to identify and measure extrasolar giant planets
\citep{2002AcA....52..317U}.
\\
\\
4. {\it To find rare variable sources.} The massive variability dataset and
multi-colour atlas will allow us to search for CVs (novae, dwarf novae) and
other eruptive variables (e.g., RS~CVn), eclipsing binary RR~Lyrae,
pre-HB/post He-flash stars \citep{2008AA...489.1201S}, eclipsing binary
red-clump giants, luminous blue variables (LBVs), FU~Ori protostars undergoing
unstable accretion, and asymptotic giant branch (AGB) stars at the stage of
unstable shell burning. The Murchison Widefield Array
\citep{2009arXiv0903.1828L} will soon open up the field of surveys of radio
transients in the southern hemisphere, likely making infrared variability
surveys of the southern part of the Galaxy interesting for interpreting their
results. The Galactic center region contains numerous high energy sources
without counterparts in the optical/infrared band, with continuing discoveries
coming both from space-based X-ray and $\gamma-$ray observatories and from
ground-based Cerenkov arrays. The most sensitive of these observatories, the
Chandra X-ray Observatory, has discovered almost 10$^4$ X-ray sources within a
$2\times0.8$~deg$^2$ region near the Galactic center
\citep{2009ApJS..181..110M}. The Chandra Galactic Bulge Survey
\citep{2008ATel.1575....1B} contains about 10$^3$ X-ray sources spread over a
larger, slightly less crowded region, and already has deep optical follow-up.
These objects are often extincted enough that they cannot be observed
in the optical, but there are also often multiple infrared counterparts found
even in moderately deep infrared images within the 1'' or smaller Chandra
error regions \citep[e.g.,][]{2007MNRAS.380.1511G}.  For $\gamma$-ray sources,
such as those which have been found with INTEGRAL
\citep[e.g.,][]{2006AIPC..840...30K, 2007ApJS..170..175B} and HESS
\citep{2007AA...469L...1A, 2008AIPC.1085..372C}, the point source localization
is even poorer -- typically several arcminutes. Fortunately, X-ray binaries
are usually variable, and their variability in infrared and X-rays is, for the
most part, well correlated \citep[e.g.,][]{2006MNRAS.371.1334R}.  As a result
an infrared variability survey of the inner Galaxy can be combined with the
ongoing monitoring in X-rays and $\gamma$-rays from missions like Swift and
MAXI, should give the opportunity to search narrow down the list of candidate
infrared counterparts through searches for IR emitters with variability well
correlated with that in the high energy bands.
\\
\\
5. {\it To search for microlensing event}, especially reddened events, short 
time-scale events, and high magnification events in obscured high-density
fields.  The spatial dependence of the microlensing optical depth $\tau$ at
near-infrared wavelengths has been modelled by \cite{2009MNRAS.396.1202K}
using synthetic population synthesis models of the Galactic disk and bulge.
The spatial variation of $\tau$ can probe directly the mass distribution
contained in the inner regions. Unfortunately, current optical microlensing
searches do not cover the whole bulge or the plane due to the prevalence of
dust. In particular, they miss the inner regions of the bulge where the
optical depth is higher, providing poor constraints to current models
\citep{2002MNRAS.330..591B,2004ApJ...601L.155B}. At near-IR wavelengths a map
of microlensing optical depth for the whole bulge can be made, allowing a
search for asymmetries in the spatial distribution of $\tau$. The strength of
the microlensing asymmetry is a function of the orientation of the inner bar
as well as the relative contributions of the bulge and disk to the
microlensing rate. A near-IR microlensing sample can therefore provide an
important additional lever on the 3D geometry of the inner Galaxy. In
addition, we expect to detect microlensing of stars in the Sgr dSph galaxy
\cite[e.g.,][]{2005ApJ...631..879P}.
   \\ 
   \\ 
6. {\it To monitor variability around the Galactic center}: an area of
1.5~deg$^{2}$ around the Galactic center including the 180~pc Nuclear Ring
\citep{2002AA...393..115M} will be the most frequently monitored field, over a
total of 200 epochs spanning five years. Expected variability due to
high-magnification microlensing, or flares due to black-hole accretion, can
occur \citep[e.g.,][]{2001ApJ...563..793C}. Black-hole flares easily reach
$K_{\rm s}$$\sim$16~mag, with a typical duration of 10--30~min. The expected
flare rate is 2--6 per day \citep{2003Natur.425..934G,2005ApJ...628..246E}, in
addition to a longer-timescale variation predicted by accretion simulations
\citep{2006MNRAS.366..358C}.  We also expect some Wolf-Rayet variability in
the population of massive stars and clusters in this region, and we will
search for eclipsing WR stars. In addition, we will be able to identify the
counterparts of high-energy ($\gamma$-, and X-ray) sources: accreting black
holes, microquasars, binary-pulsar companions, low mass X-ray binaries
(LMXBs), and high mass X-ray binaries (HMXBs) \citep[e.g.,
][]{1998Natur.392..673M,2008MNRAS.391..136L}. In particular, this survey may
reveal the as yet undetected counterparts of the most luminous, persistent
hard X-ray/jet sources in the Galactic center region, 1E~1740.7-2942
\citep{1992Natur.358..215M} and GRS~1758-258 \citep{1992ApJ...401L..15R}. Our
survey will allow us to identify and monitor the counterparts of several
variable high-energy sources, and perhaps in some cases to determine their
orbital periods. 
\\
\\
7. {\it To search for new star clusters of different ages and identify their
variable star members}, such as Cepheids, semiregular variables, W~UMa-, and
$\delta$~Sct-type stars. The asymmetric distribution of the known globulars in
the Galactic center region hints at the presence of additional, as yet
undiscovered objects \citep{2005AA...442..195I, 2008AA...489..583K}. Our team
members have already carried out successful campaigns searching for new
clusters in the 2MASS Point Source Catalogue
\citep{2002AA...394L...1I,2003AA...411...83B,2003AA...400..533D}. Note that
2MASS, with a $K_{\rm s,lim} = 14.5$~mag in the bulge, discovered hundreds of
open-cluster candidates, as well as two new globular clusters. Because we will
reach $3-4$~mag deeper, we expect to find many new clusters.

We will build a homogeneous sample of at least 300 open clusters in the
direction of the Galactic center, with accurately derived fundamental
parameters. A sample thus obtained will be useful to investigate the structure
of the Galactic disk in directions intercepting the bulge. This will represent
a major improvement over our current knowledge in this area, since only 20
clusters have so far been studied in detail in this region of the sky,
according to the last update (ver.~2.10, 2009) of Dias' catalogue of Galactic
Open Clusters \citep{2002AA...389..871D}. We will carry out a census of the
Milky Way open clusters projected onto the central parts of the Galaxy. This
will allow us to establish the fraction of star clusters compared to
statistical fluctuations of the dense stellar field in those directions, as
well as the cluster-formation efficiency relative to field stars. We can also
estimate the role of disruption effects on timescales $<10$~Myr (particularly
useful for open clusters).  We will derive the physical parameters: angular
sizes, radial velocities, reddening, distances, masses, and ages of these
clusters. Reliable fundamental parameters of unstudied open clusters are
important both to disk studies and to constrain the theories of
molecular-cloud fragmentation, star formation, as well as stellar and
dynamical evolution. We will trace the structure of the Galactic disk. Recent
studies of the disk structure based on open clusters are complete up to only 1
kpc from the Sun. We will complement and re-derive the existing kinematics
distributions such as distance of the cluster to the Galactic center $vs.$ age
distribution, open clusters age histogram, distance of the open clusters to
the Sun $vs.$ reddening, etc.  
\\ 
\\ 
8. {\it To provide complementary near-IR multi-colour information} (reddening,
temperatures, luminosities) and time coverage to the following past and
on-going surveys: GLIMPSE-II, VPHAS$+$, MACHO, OGLE, EROS, MOA, Pan-STARRS1, and
PLANET. Near-IR photometry is important for the events discovered by
microlensing surveys. For old previously detected events or new ones, the VVV
survey will give field reddening, a baseline colour, and a magnitude that can
immediately be translated into temperature and luminosity for the source
star. Characterization of the source is essential for refining the
microlensing light-curve parameters and the physical lens properties
\citep[e.g.,][]{2006Natur.439..437B}.  
\\ 
\\ 
9. {\it To find variable stars in  the Sgr dSph galaxy}: Figure\,\ref{alcock}
shows that the Sgr dSPh RR~Lyrae are well within reach and can be readily
identified.  RR~Lyrae would give the 3-D structure of the Sgr
dSph \citep[e.g.,][]{1996ApJ...458L..17A, 1998ApJ...492..190A}.  To
measure the depth and the tilt of the Sgr dSph along the line of sight, mean
RR~Lyrae magnitudes with an accuracy better than 0.01~mag are necessary.
This corresponds to the tidal radius of M54, supposedly the core
of the Sgr dSph galaxy \citep[][]{2004MNRAS.353..874M}. We will also
detect and measure carbon stars, semiregulars, and eclipsing binary members of
the Sgr dwarf galaxy (type II Cepheids are expected in the bulge, but
classical Cepheids are not; since these would be found in the disk instead).
\\ 
\\ 
10. {\it To identify high-proper motion objects and background Quasi-Stellar
Objects (QSOs)}: this goal links the -- seemingly unrelated -- intrinsically
faintest and brightest objects in the Universe. On the faint end, we would use
proper motions to find nearby late M-type stars, brown dwarfs (L and T types),
and high-velocity halo stars. The proper motions of the most interesting
low-mass objects will be determined using UKIDSS, DENIS, and 2MASS to extend
the time baseline, and to search for objects with smaller proper motions. With
an astrometric accuracy of $\sim 10$~mas, comparable to UKIDSS
\citep[e.g.,][]{2007MNRAS.379.1423L,2009MNRAS.394..857D}, or slightly lower in
the most crowded regions, we will be able to determine this proper motions
\citep{2008MNRAS.391..136L}. Scaling the results of the UKIDSS proper motion
studies we expect an accuracy of $\lesssim15$~miliarcs yr$^{-1}$. This is
beyond the minimum value to distinguish bulge and disk populations, since the
proper motion difference between them is $\sim 6$ miliarcs yr$^{-1}$ in the
galactic center region \citep{2008ApJ...684.1110C}.

On the intrinsically bright end, variability would also allow us to identify
background quasars, providing an extragalactic reference scale for future
proper motions \citep[e.g., ][]{2005AJ....130...95P}. QSOs have a relatively
broad colour range depending on their redshift, and their intrinsic
variability increases monotonically with increasing time lags
\citep{2005AJ....129..615D}. Their amplitudes should be $> 0.2$~mag in the
near-IR \citep{2002ApJS..141...45E}. We estimate that we will find $>$~500
active galactic nuclei (AGNs), assuming a surface density of $2$~deg$^{-2}$
with $K_s < 15.5$~mag \citep{2005AA...440L...5L} in the regions above and
below the disk where $A_{K} < 0.5$~mag.  
\\ 
\\ 
11. {\it To identify pre-main-sequence (pre-MS) clusters and associations
through variability}: neither the duration of pre-MS evolution as a function
of mass nor the duration of active star formation in molecular clouds are well
established. IR studies \citep[e.g.][]{2001AJ....121.3160C} have shown that a
large fraction of all pre-MS stars are variable over a wide range in absolute
magnitude, therefore, the identification of young stellar objects (YSOs) via a
variable survey is a suitable method to detect YSOs due to its independence on
color. VVV will be capable of picking out loose associations of pre-MS stars
in crowded fields long after the molecular cloud has dispersed and the cluster
has become unbound. Through careful analysis of the detection rate of such
dispersed pre-MS populations and comparison with the detection rate of pre-MS
accretion disks with Spitzer and WISE it will be possible to determine the
relative durations of the phases with and without a disk. Two such rare pre-MS
variables, KH 15D and V1648 Ori, have been described by
\cite{2002PASP..114.1167H, 2005ApJ...632L.139K} and
\cite{2004ApJ...606L.119R}.

\section{Conclusions}

We have described our near-IR public survey of the inner Milky Way bulge and
disk. The VVV is a survey to be carried out with VISTA at Paranal Observatory
between 2010 and 2014. It will map repeatedly most of the Milky Way bulge, as
well as the inner southern disk, covering a total area of about 520~deg$^{2}$
containing about $\sim 10^{9}$ point sources, 33 known globular clusters and
more than three hundred known open clusters. The main survey products will be
a $Z\,Y\,J\,H\,K_{\rm s}$ atlas of the Milky Way bulge and inner disk, and
catalogues of variable point sources and high-proper-motion objects. The
multi-epoch $K_s$-band photometry will allow the identification and phasing of
periodic variable stars, as well as microlensing events and planetary
transits. We will unveil the 3-D structure of the inner bulge and disk of the
Milky Way using well understood distance indicators such as RR~Lyrae stars,
Cepheids, and red-clump giants. The survey is expected to detect more than 100
star forming regions and pre-MS clusters, benefiting from sensibility to
variable pre-MS stars in regions where the molecular cloud has been
dispersed. This large sample will permit measurement of the duration of the
pre-MS evolution as a fuction of mass. It will also allow investigation of the
effect of environment on the outcome of the star formation process. The VISTA
observations will be combined with data from MACHO, OGLE, EROS, 2MASS, DENIS,
HST, Spitzer, Chandra, INTEGRAL, AKARI, WISE, Fermi LAT, XMM-Newton and in the
future GAIA and ALMA for a complete understanding of the variable star
sources in the inner Milky Way.

\section*{Acknowledgements}

We would like to thank Sebasti\'an Ram\'{i}rez-Alegr\'{i}a for providing us
with NTT/SOFI data showed in the bottom panel of Fig.\,\ref{tr113}. This work
is supported by MIDEPLAN's Programa Iniciativa Cient\'{i}fica Milenio through
grant P07-021-F, awarded to The Milky Way Millennium Nucleus; by the BASAL
Center for Astrophysics and Associated Technologies PFB-06; by the FONDAP
Center for Astrophysics N. 15010003; and by FONDECYT N. 1090213 and
1071002. R.~H.~Barb\'a, G.~Gunthardt and M.~Soto acknowledge support from
CONICYT through Gemini Project N. 32080001. J.~Borissova and R.~Kurtev
acknowledge support from FONDECYT N. 1080086 and 1080154. G.~Pignata
acknowledges support from the Millenium Center for Supernova Science through
MIDEPLAN grant P06-045-F and ComitMixto ESO-Gobierno de Chile.

\vskip 0.5em
\small
$^{1}$
Departamento de Astronom\'{\i}a y Astrof\'{\i}sica, Pontificia
  Universida Cat\'{o}lica de Chile, Av.~Vicu\~na Mackenna 4860, Casilla 306,
  Santiago 22, Chile
\\
$^{2}$ 
Vatican Observatory, Vatican City State V-00120, Italy
\\
$^{3}$ 
Centre for Astrophysics Research, Science and Technology
  Research Institute, University of Hertfordshire, Hatfield AL10 9AB, UK
\\
$^{4}$ 
Astronomy Unit, School of Mathematical Sciences, Queen Mary,
  University of London, Mile End Road, London, E1 4NS, UK
\\
$^{5}$ 
Nicolaus Copernicus Astronomical Center, ul. Bartycka 18,
 00-716 Warszawa, Poland
\\
$^{6}$ 
Observatorio Astron\'omico de C\'ordoba, Universidad Nacional
  de C\'ordoba, Laprida 854, 5000 C\'ordoba, Argentina
\\
$^{7}$ 
European Southern Observatory, Av. Alonso de C\'ordova 3107,
  Casilla 19, Santiago 19001, Chile
\\
$^{8}$ 
Consejo Nacional de Investigaciones Cient\'{i}ficas y
  T\'ecnicas, Av. Rivadavia 1917 - CP C1033AAJ - Buenos Aires, Argentina
\\
$^{9}$ 
Department of Astronomy, University of Michigan, Ann Arbor, MI
  48109-1090, USA
\\
$^{10}$ 
Departamento de F\'{i}sica, Universidad de La Serena,
  Benavente 980, La Serena, Chile
\\
$^{11}$ 
Department of Astronomy, University of Florida, 211 Bryant Space
 Science Center P.O. Box 112055, Gainesville, FL, 32611-2055, USA
\\
\\
$^{12}$
Universidade de S\~ao Paulo, IAG, Rua do Mat\~ao 1226, Cidade
 Universit\'aria, S\~ao Paulo 05508-900, Brazil
\\
$^{13}$ 
Space Telescope Science Institute, 3700 San Martin Drive,
  Baltimore, MD 21218, USA
\\
$^{14}$ 
Universidade Federal do Rio Grande do Sul, IF, CP 15051, Porto
  Alegre 91501-970, RS, Brazil
\\
$^{15}$ 
Departamento de F\'{i}sica y Astronom\'{i}a, Facultad de
  Ciencias, Universidad de Valpara\'{i}so, Ave. Gran Breta\~na 1111, Playa
  Ancha, Casilla 5030, Valpara\'iso, Chile
\\
$^{16}$ 
Departamento de Astronom\'{i}a, Universidad de Chile, Casilla
  36-D, Santiago, Chile
\\
$^{17}$ 
Institute for Astronomy, The University of Edinburgh, Royal
 Observatory, Blackford Hill, Edinburgh EH9 3HJ, UK
\\
$^{18}$ 
The Department of Physics and Astronomy, University of
  Sheffield, Hick Building, Hounsfield Road, Sheffield, S3 7RH, UK
\\
$^{19}$
 National Astronomical Observatories, Chinese Academy of
  Sciences, 20A Datun Road, Chaoyang District, Beijing 100021, China
\\
$^{20}$ 
Konkoly Observatory of the Hungarian Academy of Sciences,
H-1525 Budapest, PO Box 67, Hungary
\\
$^{21}$ 
Astrophysics Group, Imperial College London, Blackett
  Laboratory, Prince Consort Road, London SW7 2AZ, UK
\\
$^{22}$ 
Facultad de Ciencias Astron\'omicas y Geof\'isicas,
  Universidad Nacional de La Plata, and Instituto de Astrof\'{i}sica La Plata,
  Paseo del Bosque S/N, B1900FWA, La Plata, Argentina
\\
$^{23}$
Departmento de Astronom\'{i}a, Universidad de Concepci\'on,
  Casilla 160-C, Concepci\'on, Chile
\\
$^{24}$
Max Planck Institute for Astronomy, K\"{o}nigstuhl 17, 69117
 Heidelberg, Germany
\\
$^{25}$
European Southern Observatory, Karl-Schwarzschild-Strasse 2,
  D-85748 Garching, Germany
\\
$^{26}$
Institute of Astronomy, University of Cambridge, Madingley Road,
 Cambridge CB3 0HA, UK
\\
$^{27}$
Jodrell Bank Centre for Astrophysics, The University of
  Manchester, Oxford Road, Manchester M13 9PL, UK
\\
$^{28}$
Instituto de Astrof\'isica de Canarias, V\'{i}a L\'actea s/n,
  E38205 - La Laguna (Tenerife), Spain
\\
$^{29}$
School of Physics and Astronomy, University of Southampton,
  Highfield, Southampton, SO17 1BJ, UK
\\
$^{30}$
Istituto di Astrofisica Spaziale e Fisica Cosmica di Bologna,
  via Gobetti 101, 40129 Bologna, Italy
\\
$^{31}$
Chester F. Carlson Centre for Imaging Science, Rochester
  Institute of Technology, 54 Lomb Memorial Drive, Rochester NY 14623, USA
\\
$^{32}$
The Astrophysics and Fundamental Physics Missions Division,
  Research and Scientific Suppport Department, Directorate of Science and
  Robotic Exploration, ESTEC, Postbus 299, 2200 AG Noordwijk, the Netherlands
\\
$^{33}$
Service d'Astrophysique - IRFU, CEA-Saclay, 91191 Gif sur
  Yvette, France
\\
$^{34}$
Instituto de Astronom\'{i}a y F\'{i}sica del Espacio, Casilla de
 Correo 67, Sucursal 28, Buenos Aires, Argentina
\\
$^{35}$
Dipartimento di Astronomia, Universit\'a di Padova, vicolo
 dell'Osservatorio 3, 35122 Padova, Italy
\\
$^{36}$
Departamento de Ciencias Fisicas, Universidad Andres Bello,
 Av. Rep\'ublica 252, Santiago, Chile 
\\
$^{37}$
Department of Physics \& Astronomy, University of Leicester,
 University Road, Leicester, LE1 7RH, UK
\\
$^{38}$
Hartebeesthoek Radio Astronomy Observatory, PO Box 443,
  Krugersdorp 1740, South Africa
\\
$^{39}$
The University of Kent, Canterbury, Kent, CT2 7NZ, UK
\\
$^{40}$
Division of Optical and Infrared Astronomy, National
  Astronomical Observatory of Japan 2-21-1 Osawa, Mitaka, Tokyo, 181-8588,
  Japan

\end{document}